\documentclass[showpacs,aps,prb,twocolumn,superscriptaddress]{revtex4}
\usepackage{graphicx} 
\usepackage{dcolumn}
\usepackage{bm}
\usepackage{amssymb,amsmath}
\usepackage{epstopdf}

\begin{document}
\title{Dephasing of G-band Phonons in Single-Wall Carbon Nanotubes \\Probed via Impulsive Stimulated Raman Scattering}

\normalsize
\author{J.-H.~Kim}
\affiliation{Department of Physics, Chungnam National University, Daejeon 305-764, Korea}
\affiliation{Department of Electrical and Computer Engineering, Rice University, Houston, Texas 77005, USA}

\author{K.-J.~Yee}
\email[]{kyee@cnu.ac.kr}
\thanks{corresponding author.}
\affiliation{Department of Physics, Chungnam National University, Daejeon 305-764, Korea}

\author{Y.-S.~Lim}
\affiliation{Department of Nano Science and Mechanical Engineering and Nanotechnology Research Center, Konkuk University, Chungju, Chungbuk 380-701, Korea}

\author{L.~G.~Booshehri}
\author{E.~H.~H\'{a}roz}
\affiliation{Department of Electrical and Computer Engineering, Rice University, Houston, Texas 77005, USA}

\author{J.~Kono}
\email[]{kono@rice.edu}
\thanks{corresponding author.}
\affiliation{Department of Electrical and Computer Engineering, Rice University, Houston, Texas 77005, USA}
\affiliation{Department of Physics and Astronomy, Rice University, Houston, Texas 77005, USA}

\date{\today}

\begin{abstract}
We have studied the coherent dynamics of G-band phonons in single-wall carbon nanotubes through impulsive stimulated Stokes and anti-Stokes Raman scattering.  The probe energy dependence of phonon amplitude as well as preferential occurrence between Stokes and anti-Stokes components in response to chirped-pulse excitation are well explained within our model.  The temperature dependence of the observed dephasing rate clearly exhibits a thermally-activated component, with an activation energy that coincides with the frequency of the radial breathing mode (RBM). This fact provides a clear picture for the dephasing of G-band phonons by random frequency modulation via interaction with the RBM through anharmonicity.
\end{abstract}


\maketitle


Optical phonons often play major roles in dynamical phenomena in solids.  Due to their strong coupling with electrons, they are behind virtually all energy and phase relaxation processes that occur in the presence of high electric fields and/or non-equilibrium carrier distributions.  Recently, much attention has been paid to non-perturbative electron-optical-phonon coupling in single-wall carbon nanotubes (SWCNTs).\cite{DubayetAl02PRL,ConnetableetAl05PRL,FoaTorresRoche06PRL}  Such strong coupling is believed to be responsible for the current saturation behaviors observed in high-field electronic transport~\cite{Yaoetal00PRL,JaveyetAl04PRL} as well as for the appearance of a broad and red-shifted Raman feature due to Kohn anomalies.\cite{PiscanecetAl07PRB}  In both cases, dynamical quantities of optical phonons such as lifetimes and dephasing times are the key parameters that characterize the processes.

Raman scattering spectroscopy has been an indispensable tool for characterizing electronic structure of carbon-based nanomaterials such as SWCNTs.\cite{DresselhausetAl10NL}  While an extensive literature exists on CW Raman studies, time-domain vibrational measurements directly probing lattice dynamics in SWCNTs have only recently begun.\cite{LimetAl06NL,GambettaetAl06NP,KatoetAl08NL,SongetAl08PRL,KimetAl09PRL,LueretAl09PRL,LeeetAl10PRB}  Real-time observations of coherent oscillations have been made of both the low-frequency (100-300~cm$^{-1}$) radial-breathing mode (RBM) and high-frequency (1550-1600~cm$^{-1}$) phonon mode, the so-called G-band, providing important insight into chirality dependence~\cite{LimetAl06NL}, generation mechanisms,\cite{KimetAl09PRL,SandersetAl09PRB} and electron-phonon coupling strength.\cite{LueretAl09PRL}  A population lifetime of 1.1~$\pm$~0.2~ps was also measured for optical phonons in SWCNTs at room temperature using time-resolved anti-Stokes Raman spectroscopy.\cite{SongetAl08PRL}  However, exactly how optical phonons decay in energy and phase in SWCNTs is still an open question.

Here, we report results of coherent phonon spectroscopy of SWCNTs using spectrally-resolved and temperature-dependent ultrafast pump-probe spectroscopy.  A pump pulse initiates coherent lattice vibrations, and then a delayed, spectrally broad probe pulse is incident on the sample, where it induces additional lattice vibrations through impulsive stimulated Raman scattering.  The probe photon-energy dependence of phonon amplitude for both transformed-limited and chirped pulses can be well explained within our model based on impulsive stimulated Stokes Raman scattering (SSRS) and anti-Stokes Raman scattering (SARS).\cite{YanetAl85JCP}  The temperature dependence of the observed dephasing rate clearly exhibits a thermally-activated component, indicating the presence of the `exchange-modulation' mechanism.\cite{HarrisetAl77PRL}  Namely, the G-band phonon mode dephases via anharmonicity-induced coupling with a lower-frequency mode.  Our quantitative analysis provides evidence that the lower-frequency mode responsible for the decay of G-band phonons is the RBM.

A micelle-encapsulated SWNT suspension, where High pressure gas phase decomcarbon monoxide process (HiPco) nanotubes were individually dispersed in a 1\% sodium cholate (weight/volume) solution in D$_{2}$O~\cite{OconnelletAl02Science} and a film containing isolated HiPco nanotubes~\cite{KimetAl07AM} were used in this study. Note that we used the same batch of nanotubes here, compared to that in our previous paper.\cite{LimetAl06NL} Using 12~fs long pulse from a Ti:Sapphire laser, we performed pump-probe measurements in a transmission geometry.  The center photon energy of the laser spectrum was $\sim$1.55~eV (800~nm), with a spectral bandwidth of $\sim$200~meV (100~nm), which is comparable to the G-mode vibrational energy (197~meV) corresponding to the longitudinal optical (LO) phonons in semiconducting SWCNTs.  The probe beam was spectrally filtered after the sample and before the detector, by using a series of band-pass filters with a 10~nm pass band-width centered at various wavelengths.  The pump pulse fluence was 0.14~mJ/cm$^{2}$, and the probe one was a tenth of the pump one.  Pump-induced and spectrally-resolved transmission or scattering modulations were measured as a function of the time delay between pump and probe pulses.

\begin{figure}
\includegraphics[scale=0.44]{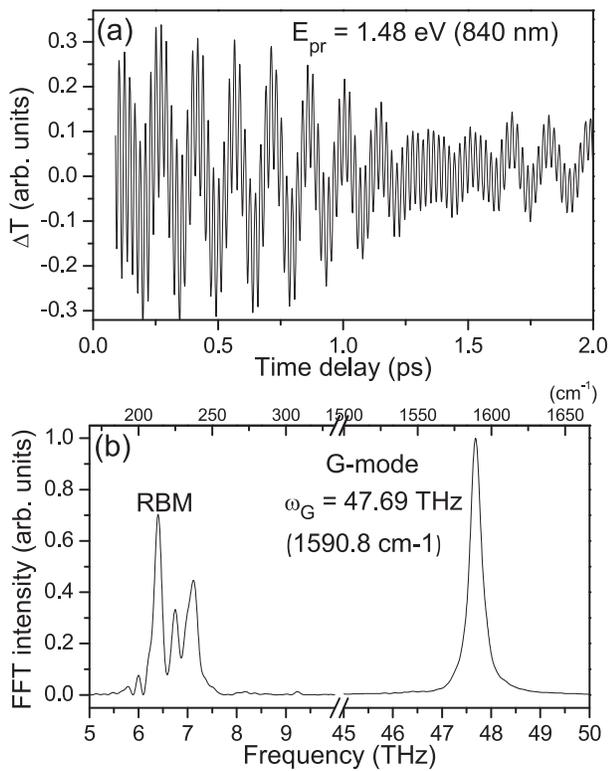}
\caption{(a) Transmitted intensity modulations due to coherent phonon oscillations in SWCNTs, which were extracted from the pump-probe signal for pump and probe energies of 1.55~eV (800~nm,~$E_{\rm pu}$) and 1.48~eV (840~nm,~$E_{\rm pr}$), respectively. (b)~Corresponding Fourier-transformed spectrum showing radial-breathing modes (RBMs) at 6.0-7.5~THz (200-250~cm$^{-1}$) and G-mode phonons at 47.69~THz (1590.8~cm$^{-1}$).}
\label{typical}
\end{figure}

Shown in Fig.~1(a) are time-domain transmission modulations of the SWNT suspension for a probe photon energy of 1.48~eV (840~nm), obtained after subtracting an exponentially decaying component that corresponds to carrier relaxation.\cite{OstojicetAl04PRL,OstojicetAl05PRL} The oscillatory signal, which originates from the coherent lattice vibrations excited by the pump pulse, consists of high-frequency and low-frequency contributions.  As is confirmed in the Fourier transform in Fig.~1(b), the low-frequency signal at $\sim$7~THz corresponds to the RBM, where the existence of multiple RBM peaks indicates that the sample contains several chiralities of SWCNTs, each having a different RBM frequency, that are resonantly excited at 840~nm.\cite{LimetAl06NL} The focus of this paper is on the high-frequency, optical G-mode phonons of SWCNTs, having a frequency of 47.69~THz (1590.8~cm$^{-1}$) seen in Fig.~1(b).

For the RBM, absorption coefficient oscillations in time can be readily understood as a result of diameter-dependent bandgap.\cite{LimetAl06NL,KimetAl09PRL,SandersetAl09PRB} On the other hand, while the G-mode phonons can also modify optical constants of SWCNTs, according to theoretical calculations,\cite{SandersetAl09PRB} absorption coefficient modulations due to G-mode phonons are expected to be $\sim$1000 times smaller than those by the RBM.  The fact that the coherent optical G-mode phonon signal is comparable to the RBM signal (shown in Fig.~1) thus indicates that a different mechanism is at work.

\begin{figure}
\includegraphics[scale=0.56]{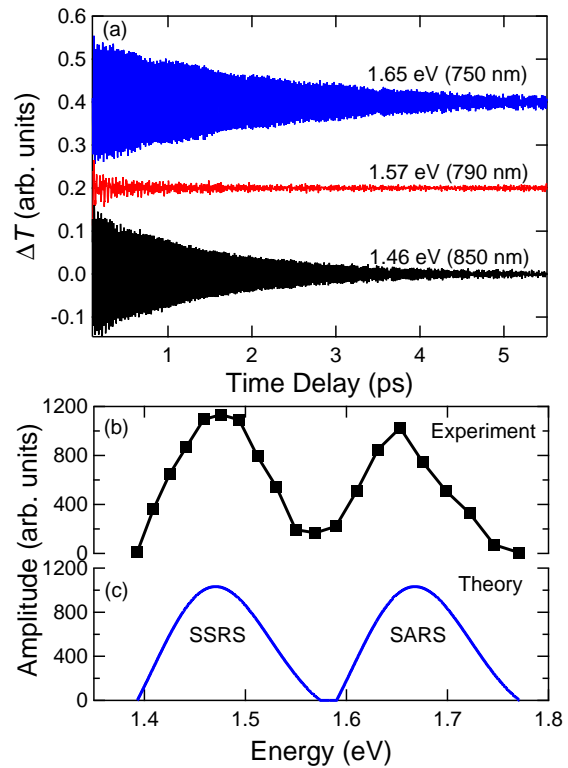}
\caption{(color online) (a) Coherent G-mode phonons dynamics measured at probe energies of 1.65~eV (750~nm), 1.57~eV (790~nm), and 1.46~eV (850~nm). The pump energy was 1.55~eV (800~nm). (b) Coherent G-mode phonon amplitude vs.~probe energy, exhibiting two peaks at 1.47~eV and 1.66~eV, corresponding to the SSRS and the SARS processes, respectively. (c) Simulated spectral intensity for the SARS and SSRS processes obtained for a Gaussian laser spectrum centered at 1.55~eV with a spectral width of 195~meV FWHM.}
\label{three-wavelength}
\end{figure}

As the spectral window for the probe pulse is shifted, the G-mode phonons show drastic changes with the probe wavelength both in amplitude and phase.  Figure~2(a) shows probe transmission modulations due to coherent G-mode phonons at different probe energies of 1.65~eV (750~nm), 1.57~eV (790~nm), and 1.46~eV (850~nm).  The pure G-mode phonon oscillation signal was obtained by removing the RBM frequency components from the data in Fig.~1(a).  It is interesting that the signal is almost completely suppressed when the probe energy is close to the center of the laser spectrum, while strong oscillations are observed at 1.65~eV and 1.46~eV, which are each separated from the center energy by roughly a half of the G-mode phonon energy ($\sim$100~meV).  Additionally, the G-mode amplitude decays monotonically with time delay for each wavelength, while there is a slight tendency that the signal at lower probe energy decays faster than that at higher energy.  Figure 2(b) shows the G-mode phonon amplitude as a function of probe energy.  The amplitude curve features two peaks with each maximum occurring near the probe energy of 1.46~eV and 1.66~eV, respectively, which are separated from each other by the G-mode phonon energy, while having a local minimum near the center energy of the laser spectrum.


\begin{figure}
\includegraphics [scale=0.43] {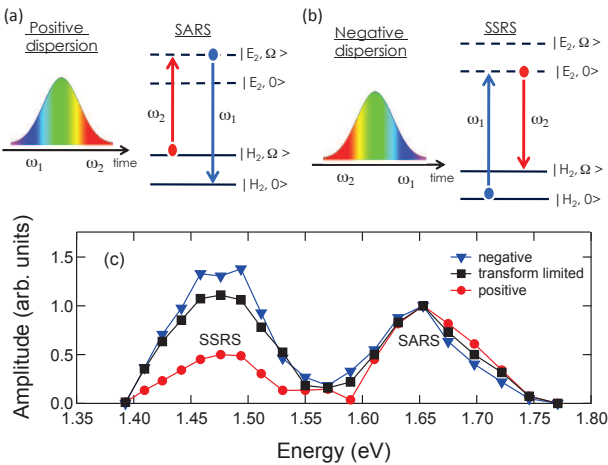}
\caption{(color online) Chirped pulse enhancing the relative amplitude of the (a) stimulated anti-Stokes Raman scattering (SARS) and (b) stimulated Stokes Raman scattering (SSRS) processes.  (c) Detected G-mode phonon amplitude vs.~probe photon energy, normalized to the amplitude at 1.65~eV, obtained with negative dispersion (blue), transform-limited (black), and positive dispersion (red) pulse chirp.}
\label{dispersion}
\end{figure}

The probe energy dependence of the G-mode phonon amplitude in Figs.~2(a) and 2(b) can be explained by taking into account impulsive SSRS and SARS processes.  The probability $P$ for the stimulated Stokes and anti-Stokes scattering to occur at a certain energy $E$ will be proportional to both the stimulating photon intensity $I(E)$ and the stimulated photon intensity $I(E -\hbar \omega_{\rm op})$ and $I(E + \hbar \omega_{\rm op})$, respectively, such that
\begin{eqnarray}
P_{\rm SSRS} (E) \propto I(E) \times I(E - \hbar \omega_{\rm op}) \times n_{\rm op} \\
P_{\rm SARS} (E) \propto I(E) \times I(E + \hbar \omega_{\rm op}) \times n_{\rm op},
\end{eqnarray}
where $\omega_{\rm op}$ is the G-mode phonon frequency and $n_{\rm op}$ is the G-mode phonon occupation number generated though the pump pulse.  Simulations of the scattering intensity for the two coherent Raman scattering processes, as depicted in Fig.~2(c), were performed assuming a Gaussian laser spectrum centered at 1.55~eV with a spectral width of 195~meV FWHM, similar to our experimental conditions.  The simulation results [Fig.~2(c)] show that SSRS (SARS) intensity will be strong on the lower (higher) energy side of the center of the laser spectrum with a peak 100~meV below (above) the center. The good agreement between the experimental data in Fig.~2(b) and the simulation in Fig.~2(c) supports the proposed detection mechanism of impulsive SSRS and SARS processes of the probe pulses. Furthermore, the photon energy of the amplitude dip changed as we tuned the center of the laser spectrum, which excludes the possibility that the probe energy dependence is related to the electronic transitions of the nanotubes.\cite{LueretAl09PRL}


For impulsive stimulated Raman scattering where two photons with energy separation by the phonon energy are involved, the time sequence between those photons influences the scattering efficiency, especially when a real electronic transition is incorporated such that the excited state can be sustained for an extended period.\cite{BardeenetAl95PRL}  If a photon with lower energy proceeds a higher energy photon, the SARS process [Fig.~3(a)] will be more efficient than the SSRS process [Fig.~3(b)], and vice versa.  The sequence of photons can be controlled by adding or subtracting dispersion by adjusting the optical path length through a prism.  The modification of the dispersion can result in chirp such that the short (long) wavelength components arrive earlier than the long (short) components for the case of negative (positive) dispersion [see Figs.~3(a) and 3(b)].  We find that the applied dispersion value of $\frac{{\rm d^2}\varphi}{{\rm d}\omega^2} = 21$~fs$^2$ modifies the ratio between the SARS contribution observed around 1.66~eV and the SSRS contribution around 1.46~eV, as demonstrated in Fig.~3(c). With positive (negative) dispersion, the SARS (SSRS) signal is stronger, in good agreement with the consideration of the order between stimulating and stimulated photons for the detection processes.  

\begin{figure}
\includegraphics[scale=0.5]{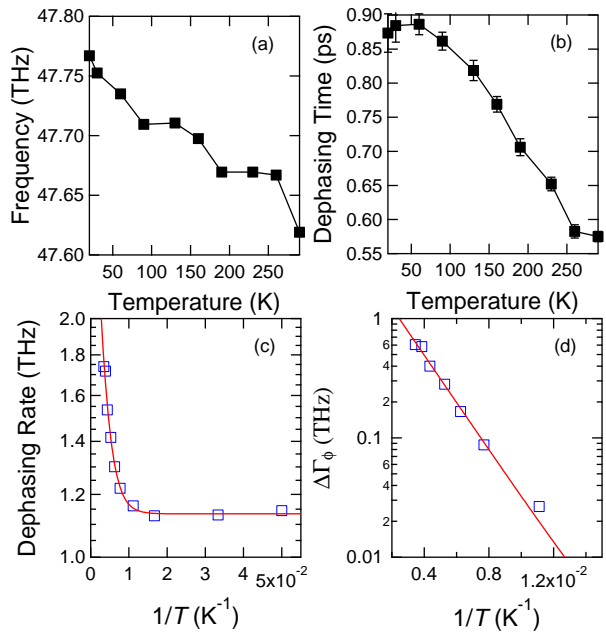}
\caption{(color online) Temperature ($T$) dependence of coherent G-mode phonons in SWCNTs measured at a probe energy of 1.65\,eV (750\,nm).  As $T$ increases, the (a)~frequency and (b)~dephasing time decrease.  (c)~Dephasing rate vs.~1/$T$.  The red trace is a theoretical fit (see text).  (d)~Thermally activated component of dephasing.  The fit (red trace) gives an activation energy of 31\,meV (= 250\,cm$^{-1}$).}
\label{temp-dep}
\end{figure}

Figures 4(a) and 4(b) show the temperature ($T$) dependent frequency and dephasing time, respectively, of coherent G-mode phonons, which were measured at pump and probe energies of 1.55\,eV (800\,nm) and 1.65\,eV (750\,nm), respectively.  The frequency decreases with increasing temperature, but the change is small, going from 47.78~THz at 20~K to 47.60~THz at 290~K, a shift of 0.18~THz (6~cm$^{-1}$).  This frequency shift is smaller than the dephasing rate (i.e., the linewidth).  On the other hand, the dephasing time $\tau_{\phi}$ changes more significantly.  It is more or less constant at low temperatures (20-50~K) but decreases rapidly above 50~K, indicating the presence of a thermally-activated dephasing mechanism.  The temperature dependence of the dephasing rate $\Gamma_{\phi} = \tau_{\phi}^{-1}$ can be well fit by $\Gamma_{\phi}(T) = \Gamma_0 + \Delta\Gamma_{\phi}$d,\cite{HarrisetAl77PRL} where $\Gamma_0$ is a constant and $\Delta\Gamma_{\phi} \propto \exp(-\Delta E /k_{\rm B}T)$, as shown by the red trace in Fig.~4(c).  The Arrhenius plot of $\Delta\Gamma_{\phi}$ [Fig.~4(d)] further confirms this thermal activation behavior, determining the activation energy $\Delta E$ = 31~meV = 250~cm$^{-1}$. This activation energy is close to the resonant frequency of the dominant (9,4) nanotubes excited at an energy of 1.65\,eV (750\,nm) among the several RBM modes excited simultaneously.

The fact that the thermal activation energy coincides with the average RBM frequency provides a clear picture for the dephasing of G-mode phonons.  High-frequency vibration modes such as G-mode phonons are susceptible to random frequency modulations by interaction (or, more specifically, energy exchange) with lower-frequency modes such as RBMs through lattice anharmonicity.\cite{HarrisetAl77PRL}  Since the rate of energy exchange is proportional to the population of the lower-frequency mode, this interaction results in thermally-activated dephasing of the higher-frequency modes, often seen in molecular systems~\cite{HessPrasad80JCP,DeSilvestrietAl85CPL,RectorFayer98JCP} and known as the exchange-modulation mechanism.

In conclusion, we studied coherent G-mode phonons in SWNT through wavelength- and temperature-dependent ultrafast pump-probe spectroscopy.  The wavelength dependence clearly shows that the detection of coherent G-mode phonons occurs through impulsive stimulated Stokes and anti-Stokes Raman scattering of probe pulses.  The phonon amplitude was strong when the probe energy was red- or blue-shifted by a half of the phonon energy from the center of the laser spectrum, originating from the SSRS and the SARS processes, respectively.  Preferential occurrence of SSRS or SARS could be obtained by controlling the temporal sequence of the probe spectral components by adding extra dispersions.  The temperature dependence of the observed dephasing rate clearly exhibits a thermally-activated component, with an activation energy that coincides with the frequency of the radial breathing mode.  Our work thus provides direct time-domain evidence for G-mode phonon dephasing in SWCNTs by interaction with the lower-frequency radial breathing mode through anharmonicity.


This work was supported by the National Research Foundation of Korea (NRF) Grants funded by the Korean Government (Basic Science Research Program: 2012-000968, 2010-0022691, SRC: 2008-0062254), the National Science Foundation (OISE-0968405), the DOE/BES through Grant No.~DEFG02-06ER46308, and the Robert A.~Welch Foundation (C-1509). We thank S.~M.~Choi for providing the film sample.


\end{document}